\documentclass[a4paper,12pt]{article}
\usepackage{amssymb}
\usepackage{amsmath}
\usepackage{amstext}
\usepackage{amsfonts}
\usepackage{epsfig}
\usepackage{dcolumn}
\usepackage{rotating}
\usepackage{color}
\usepackage{comment}
\usepackage{hyperref}

\def\XXint#1#2#3{{\setbox0=\hbox{$#1{#2#3}{\int}$ }
\vcenter{\hbox{$#2#3$ }}\kern-.56\wd0}}

\newcommand*\xbar[1]{%
  \hbox{%
    \vbox{%
      \hrule height 0.5pt 
      \kern0.5ex
      \hbox{%
        \kern-0.1em
        \ensuremath{#1}%
        \kern-0.1em
      }%
    }%
  }%
}

\definecolor{rosso}{cmyk}{0,1,1,0.4}
\definecolor{rossos}{cmyk}{0,1,1,0.55}
\definecolor{rossoc}{cmyk}{0,1,1,0.2}
\definecolor{blu}{cmyk}{1,1,0,0.3}
\definecolor{blus}{cmyk}{1,1,0,0.6}
\definecolor{bluc}{cmyk}{1,1,0,0.1}
\definecolor{verde}{cmyk}{0.92,0,0.59,0.25}
\definecolor{verdec}{cmyk}{0.92,0,0.59,0.15}
\definecolor{verdes}{cmyk}{0.92,0,0.59,0.7}

\newcommand{\ba}{\begin{eqnarray}}
\newcommand{\ea}{\end{eqnarray}}
\newcommand{\be}{\begin{equation}}
\newcommand{\ee}{\end{equation}}
\newcommand{\bi}{\begin{itemize}}
\newcommand{\ei}{\end{itemize}}
\newcommand{\al}{\alpha}
\newcommand{\bt}{\beta}

\newcommand{\da}{\delta}
\newcommand{\la}{\lambda}

\newcommand{\sa}{\sigma}

\newcommand{\cF}{{\cal F}}

\newcommand{\p}{\partial}

\newcommand{\Ra}{\Rightarrow}

\newcommand{\LF}{\left(}
\newcommand{\RF}{\right)}
\newcommand{\LT}{\left[}
\newcommand{\RT}{\right]}

\newcommand{\mt}{\mathtt}

\newcommand{\non}{\nonumber\\}

\usepackage[normalem]{ulem}

\begin{document}

\title{Slavnov Identities for Infinite Derivative Theory of Gravity}
\author{Spyridon Talaganis \\ \\
 {\it Consortium for Fundamental Physics,} \\
{\it Lancaster University, Lancaster,} \\
{\it LA$1$ $4$YB, United Kingdom.}\\
\begin{footnotesize}\textit{E-mail}:  s.talaganis@lancaster.ac.uk \end{footnotesize}}

\date{}

\maketitle

\begin{abstract}
In this paper, we shall consider an infinite-derivative theory of gravity, with a view to making it renormalisable. First, we derive the modified superficial degree of divergence. Next, we establish that the theory is invariant under BRS (Becchi-Rouet-Stora) transformations. Consequently, we show that the corresponding Slavnov identities hold for the theory. Finally, we summarise our results and discuss renormalisability.
\end{abstract}

\tableofcontents

\section{Introduction}
\numberwithin{equation}{section}

Renormalisability plays a very important role in establishing a successful theory of quantum gravity. Pure gravity is a non-renormalisable theory~\cite{tHooft:1974toh}.
Moreover, higher-derivative gravitational theories are typically plagued by ghosts.
In the seminal paper by Stelle~\cite{Stelle:1977ry,Stelle:1976gc},
it was shown that fourth-derivative gravity is renormalisable, albeit at the expense of including the Weyl ghost.

Non-locality~\cite{Moffat-qg,Tomboulis,Modesto,Kuzmin:1989sp} is a promising framework with connections to string theory~\cite{Eliezer:1989cr} within which the quest for a satisfactory theory of quantum gravity can be realised. Using spin projector operators~\cite{VanNieuwenhuizen:1973fi} for the physical degrees of freedom for the graviton, ghosts can be avoided in the context of \textit{infinite-derivative} theories of gravity~\cite{Biswas:2011ar,Biswas:2013kla}.
One can ensure that no new propagating degrees of freedom are introduced in the modified graviton propagator, apart from the massless graviton. 
Therefore, we would like to propose a \textit{ghost-free, nonlocal}, \textit{infinite-derivative} gravitational action and investigate its renormalisability.

 A ghost-free gravitational theory was presented in Refs.~\cite{Biswas:2011ar,Biswas:2013kla}.
In the action derived by Biswas, Gerwick, Koivisto and Mazumdar (BGKM), the graviton propagator is modulated by the exponential of an \textit{entire function} $a(-k^2) = e^{k^2/M^2}$~\cite{Biswas:2005qr}, 
\be
\Pi(k^2) = \frac{1}{k^2a(-k^2)}\LF {\cal P}^2 - \frac{1}{2} {\cal P}_s ^0  \RF=\frac{1}{a(-k^2)} \Pi_{GR}\,,
\ee
resulting in no poles in the propagator apart from the massless graviton. In the infrared (IR), the physical graviton propagator of General Relativity (GR) is recovered, as it should be the case.

Within the context of an infinite-derivative scalar toy model, UV finiteness of Feynman diagrams has been argued (by the use of dressed vertices and dressed propagators)~\cite{Biswas:2014tua,Talaganis:2014ida,Talaganis:2017tnr}. Moreover, within that context, the external momentum dependence of scattering diagrams is convergent for large external momenta~\cite{Talaganis:2016ovm}.
Again in that context, in a system of $n$ particles, the mass scale of non-locality decreases as the number of particles $n$ increases~\cite{Talaganis:2017dqy}.
Furthermore, in Ref.~\cite{Tal}, the Hamiltonian for an infinite-derivative theory of gravity was derived and the corresponding number of degrees of freedom was computed.

The outline of this paper is as follows. First, in section~\ref{sec:BGKM}, we write down the quantised BGKM action. Then, in section~\ref{sec:sdg}, we derive the modified superficial degree of divergence for the action.
In section~\ref{grav}, we show that the action is invariant under BRS transformations.
In section~\ref{slav}, we demonstrate that the corresponding Slavnov identities hold for the action. Finally, in section~\ref{sec:concl}, we discuss renormalisability.

\section{Gravitational action}
\numberwithin{equation}{section}
\label{sec:BGKM}

If we consider metric fluctuations around the Minkowski background,
\be
g _ {\mu \nu} = \eta _ {\mu \nu} + h _ {\mu \nu}\,,
\ee
we can define the quantum theory in harmonic gauge,
\be
\p _ {\nu} \bar{h} ^ {\mu \nu} = 0\,,
\ee
where
\be
\bar{h} _ {\mu \nu} = h _ {\mu \nu} - \frac{1}{2} \eta _ {\mu \nu} h \Ra \bar{h} ^ {\mu \nu} = h ^ {\mu \nu}-\frac{1}{2}\eta^{\mu \nu}h \, .
\ee
$\bar{h}_{\mu \nu}$ is called the trace-reverse of $h_{\mu \nu}$ since $\bar{h} = \eta^{\mu \nu} \bar{h}_{\mu \nu} =-h$, where $h = \eta^{\mu \nu} h_{\mu \nu}$. Note that $h ^ {\mu \nu} = \eta ^ {\mu \rho} \eta ^ {\nu \sigma} h _ {\rho \sigma} \Ra \bar{h}^{\mu \nu}=\eta^{\mu \rho} \eta^{\nu \sigma}\bar{h}_{\rho \sigma}$. Furthermore, $h_{\mu \nu}=\bar{h}_{\mu \nu}- \frac{1}{2}\eta_{\mu \nu}\bar{h}$.

The quantised BGKM action must contain:
\begin{align}
S _ {\mt{quantised}}& = S + S _ {\mt{GF}} + S _ {\mt{ghost}}\non
& = S_{EH}+S_{Q}- \frac{1}{2 \xi} \int{d ^ {4} x \, F _ {\tau} \cF (\Box) F ^ {\tau}}+\int{d ^ {4} x \, \bar{C} _ {\tau} \vec{F} _ {\mu \nu} ^ {\tau} D _ {\alpha} ^ {\mu \nu} C ^ {\alpha}}\,.
\end{align}
where 
\begin{align}
S & = S _{EH}+S_Q \,, \\
S_{EH} & = \frac{1}{2} \int d^{4} x \, M_{P}^{2}  R \,, \\
S_{Q} &= \frac{1}{2} \int d^{4} x \, \LF R \cF_{1}(\bar{\Box})R+R_{\mu \nu} \cF_{2}(\bar{\Box})R^{\mu \nu}+R_{\mu \nu \la \sa} \cF_{3}(\bar{\Box})R^{\mu \nu \la \sa} \RF \,,
\end{align}
where $\bar{\Box}=\Box/M^{2}$. $S_{\mt{GF}}$ is the gauge-fixing term and $S_{\mt{ghost}}$ is the ghost-antighost action while $\xi$ is a finite parameter. We have that $F ^ {\tau} = \vec{F} _ {\mu \nu} ^ {\tau} h ^ {\mu \nu}$ and $\vec{F} _ {\mu \nu} ^ {\tau} = \delta _ {\mu} ^ {\tau} \vec{\p}_ {\nu} - \frac{1}{2}\da_{\sigma}^{\tau}\eta^{\sigma \rho}\eta_{\mu \nu}\vec{\p}_{\rho}$ (the arrow indicates the direction in which the derivative acts). $C ^ {\sigma}$ is the ghost field and $\bar{C} _ {\tau}$ is the antighost field; both are anticommuting. $D _ {\al}^{\mu \nu}$ is the operator generating gauge transformations in the graviton field $h^{\mu \nu}$, given an arbitrary infinitesimal vector field $\xi ^ {\alpha} (x)$ (corresponding to $x ^ {' \mu} = x ^ {\mu} - \xi ^ {\mu}$).

That is, $\delta h _ {\mu \nu} = \delta g _ {\mu \nu}= \mathcal{L} _ {\xi} g _ {\mu \nu} = \xi ^ {\rho} \partial _ {\rho} g _ {\mu \nu} + g _ {\mu \rho} \partial _ {\nu} \xi ^ {\rho} + g _ {\rho \nu} \partial _ {\mu} \xi ^ {\rho} = D _ {\mu \nu \alpha} \xi ^ {\alpha}$, where $\mathcal{L}$ is the Lie derivative and
\be
D _ {\mu \nu \alpha} \xi ^ {\alpha} = \partial _ {\mu} \xi _ {\nu} + \partial _ {\nu} \xi _ {\mu} \nonumber +  h _ {\alpha \nu} \partial _ {\mu} \xi ^ {\alpha} + h _ {\mu \alpha} \partial _ {\nu} \xi ^ {\alpha} + \xi ^ {\alpha} \partial _ {\alpha} h _ {\mu \nu}\,.
\ee
Accordingly,
\be
\label{D}
D _{\mu \nu \al} = \eta_{\al \nu} \p_{\mu} + \eta_{\mu \al}\p_{\nu} + h _{\al \nu} \p_{\mu}+ h_{\mu \al}\p_{\nu} + \p_{\al} h _{\mu \nu}\,.
\ee
We can raise indices in~\eqref{D} using the Minkowski tensors.

Moreover, the $\cF_{i}$'s, $i=1,2,3$, are analytic functions of $\Box$:
\be
\cF_{i} (\bar{\Box}) = \sum _ {n=0}^{\infty} f_{i_{n}} \bar{\Box}^{n} \,,
\ee
where the$f_{i_n}$'s are real coefficients.

If we change the gauge-fixing term to read $F^{\tau}=e^{\tau}(x)$~\cite{Stelle:1976gc}, with $e^{\tau}(x)$ an arbitrary $4$-vector function, we can smear out the gauge condition with a weighting functional. Choosing the weighting functional
\be
\omega (e ^ {\tau}) = \exp \LT i\LF-\frac{1}{2 \xi} \int{d ^ 4 x \, e _ {\tau} \cF(\Box) e ^ {\tau}}\RF\RT\,,
\ee
where $\xi$ is a finite parameter, we obtain the gauge-fixing term (see~\cite{Stelle:1976gc} for a derivation of the gauge-fixing term in non-local theories),
\be
S_{\mt{GF}}=-  \frac{1}{2 \xi} \int{d ^ {4} x \, F _ {\tau} \cF (\Box) F ^ {\tau}}\,.
\ee
Finally, the Faddeev-Popov ghost-antighost action is given by
\be
S_{\mt{ghost}} = \int{d ^ {4} x \, \bar{C} _ {\tau} \vec{F} _ {\mu \nu} ^ {\tau} D _ {\alpha} ^ {\mu \nu} C ^ {\alpha}}\,.
\ee
It should be pointed out that the Faddeev-Popov ghost fields are benign (in contrast to the Weyl ghost). 
In the next section, we shall derive the modified superficial degree of divergence.

\section{Modified superficial degree of divergence for infinite-derivative gravitational action}
\numberwithin{equation}{section}
\label{sec:sdg}

To proceed, let us introduce the following notations:
\bi
\item $n_{h}$ is the number of graviton vertices,
\item $n_{G}$ is the number of anti-ghost-graviton-ghost vertices,
\item $n_{K}$ is the number of $K$-graviton-ghost vertices,
\item $n_{L}$ is the number of $L$-ghost-ghost vertices,
\item $I_{h}$ is the number of internal graviton propagators,
\item $I_{G}$ is the number of internal ghost propagators,
\item $E_{C}$ is the number of external ghosts,
\item $E_{\bar{C}}$ is the number of external anti-ghosts.
\ei
By counting the exponential contributions of the propagators and the vertex factors, we can now obtain the superficial degree of divergence, which is given by
\be
E = n _ {h} - I _ {h}\,,
\ee
where $n _ {h}$ is the number of graviton vertices, and $I _ {h}$ is the number of internal graviton propagators.
By using the following topological relation,
\be
L = 1 + I _ {h} + I _ {G} - n _ {h} - n _ {G} - n _ {K} - n _ {L}\,,
\ee
we get
\be
E = 1 - L + I _ {G} - n _ {G} - n _ {K} - n _ {L}\,.
\ee
Employing the momentum conservation law for ghost and anti-ghost lines,
\be
2 I _ {G} - 2 n _ {G} = 2 n _ {L} + n _ {K} - E _ {C} -E _ {\bar{C}}\,,
\ee
we obtain
\be
E = 1 - L -\frac{1}{2} \left(n _ {K} + E _ {C} + E _ {\bar{C}}\right)\,.
\ee
Note that, as $n_{K}$, $E_C$ and $E_{\bar{C}}$ increase, the degree of divergence decreases. Therefore, the most divergent diagrams are those for which $n_{K}=E_C=E_{\bar{C}}=0$, \textit{i.e.}, the diagrams whose external lines are all gravitons. In this case, the degree of divergence is given by: $E = 1 - L$. For $L >1$, $E<0$ and the corresponding loop amplitudes are superficially convergent.

\section{BRS transformations for infinite-derivative gravitational action}
\numberwithin{equation}{section}
\label{grav}

The BRS (Becchi-Rouet-Stora) transformations for BGKM gravity,  appropriate for the gauge-fixing term, $S_{\mt{GF}}$, are given by
\begin{align}
\delta _ {\mt{BRS}} h ^ {\mu \nu} & = D _ {\alpha} ^ {\mu \nu} C ^ {\alpha} \delta \lambda\,, \\
\delta _ {\mt{BRS}} C ^ {\alpha}& = - \partial _ {\beta} C ^ {\alpha} C ^ {\beta} \delta \lambda\,, \\
\delta _ {\mt{BRS}} \bar{C} _ {\tau}& =-  \frac{1}{\xi} \, \cF(\Box) F _ {\tau} \delta \lambda\,,
\end{align}
where $\da \la$ is an infinitesimal anticommuting constant parameter. The BRS transformation of the gravitational field is just a gauge transformation of $h ^ {\mu \nu}$ generated by $C ^ {\alpha} \delta \lambda$; thus, gauge-invariant functionals of $h _ {\mu \nu}$, like $S$, are BRS-invariant.

\subsection{Nilpotence of the Ghost Field}

The transformation of $C ^ {\alpha}$ is nilpotent:
\begin{align}
\delta _ {BRS} \left(\partial _ {\beta} C ^ {\sigma} C ^ {\beta} \right) & =  \delta _ {BRS} \left(\partial _ {\beta} C ^ {\sigma} \right) C ^ {\beta} + \partial _ {\beta} C ^ {\sigma} \delta _ {BRS} C ^ {\beta} \nonumber \\
& =  \partial _ {\beta} \left(-  \partial _ {\lambda} C ^ {\sigma} C ^ {\lambda} \delta \lambda \right) C ^ {\beta} + \partial _ {\beta} C ^ {\sigma} \left(-  \partial _ {\lambda} C ^ {\beta} C ^ {\lambda} \delta \lambda \right) \nonumber \\
& =  -  \partial _ {\beta} \partial _ {\lambda} C ^ {\sigma} C ^ {\lambda} \delta \lambda C ^ {\beta} -  \partial _ {\lambda} C ^ {\sigma} \partial _ {\beta} C ^ {\lambda} \delta \lambda C ^ {\beta} -  \partial _ {\beta} C ^ {\sigma} \partial _ {\lambda} C ^ {\beta} C ^ {\lambda} \delta \lambda \nonumber \\
& =  0 \,,
\end{align}
where the first term vanishes by the interchange of $\beta$ and $\lambda$ and the commutativity of partial derivatives and the last two terms cancel due to the anticommutativity of $C ^ {\beta}$ and $\delta \lambda$ and relabelling of indices. We have also assumed that $\partial _ {\beta}$ and $\delta _ {BRS}$ commute.

\subsection{Nilpotence of the Graviton Field}

The transformation of $h ^ {\mu \nu}$ is also nilpotent:
\begin{align}
\delta _ {\mathrm{BRS}} \left(D _ {\alpha} ^ {\mu \nu} C ^ {\alpha} \right) & =  \delta _ {\mathrm{BRS}} \left(\partial ^ {\mu} C ^ {\nu} + \partial ^ {\nu} C ^ {\mu} - \eta ^ {\mu \nu} \partial _ {\alpha} C ^ {\alpha} +  \left(\partial _ {\alpha} C ^ {\mu} h ^ {\alpha \nu} +  \partial _ {\alpha} C ^ {\nu} h ^ {\alpha \mu} - C ^ {\alpha} \partial _ {\alpha} h ^ {\mu \nu} - \partial _ {\alpha} C ^ {\alpha} h ^ {\mu \nu}\right) \right) \nonumber \\
& =  \partial ^ {\mu} \delta _ {\mathrm{BRS}} C ^ {\nu} + \partial ^ {\nu} \delta _ {\mathrm{BRS}} C ^ {\mu} - \eta ^ {\mu \nu} \partial _ {\alpha} \delta _ {\mathrm{BRS}} C ^ {\alpha} +  (\partial _ {\alpha} \delta _ {\mathrm{BRS}} C ^ {\mu} h ^ {\alpha \nu} + \partial _ {\alpha} C ^ {\mu} \delta _ {\mathrm{BRS}} h ^ {\alpha \nu} \nonumber \\
& + \partial _ {\alpha} \delta _ {\mathrm{BRS}} C ^ {\nu} h ^ {\mu \alpha} + \partial _ {\alpha} C ^ {\nu} \delta _ {\mathrm{BRS}} h ^ {\mu \alpha} - \delta _ {\mathrm{BRS}} C ^ {\alpha} \partial _ {\alpha} h ^ {\mu \nu} - C ^ {\alpha} \partial _ {\alpha} \delta _ {\mathrm{BRS}} h ^ {\mu \nu} \nonumber \\
& - \partial _ {\alpha} \delta _ {\mathrm{BRS}} C ^ {\alpha} h ^ {\mu \nu} -\partial _ {\alpha} C ^ {\alpha} \delta _ {\mathrm{BRS}} h ^ {\mu \nu} ) \nonumber \\
& =  - C ^ {\alpha} (\partial _ {\alpha} \partial ^ {\mu} C ^ {\nu} + \partial _ {\alpha} \partial ^ {\nu} C ^ {\mu} - \eta ^ {\mu \nu} \partial _ {\alpha} \partial _ {\lambda} C ^ {\lambda} +  (\partial _ {\alpha} \partial _ {\lambda} C ^ {\mu} h ^ {\lambda \nu} + \partial _ {\lambda} C ^ {\mu} \partial _ {\alpha} h ^ {\lambda \nu} + \partial _ {\alpha} \partial _ {\lambda} C ^ {\nu} h ^ {\mu \lambda} \nonumber \\
& + \partial _ {\lambda} C ^ {\nu} \partial _ {\alpha} h ^ {\mu \lambda} )) \delta \lambda \nonumber \\
& +  \partial _ {\alpha} C ^ {\mu} (\partial ^ {\alpha} C ^ {\nu} + \partial ^ {\nu} C ^ {\alpha} - \eta ^ {\alpha \nu} \partial _ {\lambda} C ^ {\lambda} +  \left(\partial _ {\lambda} C ^ {\alpha} h ^ {\lambda \nu} +  \partial _ {\lambda} C ^ {\nu} h ^ {\alpha \lambda} \right. \nonumber \\
& \left. - C ^ {\lambda} \partial _ {\lambda} h ^ {\alpha \nu} - \partial _ {\lambda} C ^ {\lambda} h ^ {\alpha \nu}\right) ) \delta \lambda \nonumber \\
& +  \partial _ {\alpha} C ^ {\nu} (\partial ^ {\alpha} C ^ {\mu} + \partial ^ {\mu} C ^ {\alpha} - \eta ^ {\alpha \mu} \partial _ {\lambda} C ^ {\lambda} +  \left(\partial _ {\lambda} C ^ {\alpha} h ^ {\lambda \mu} +  \partial _ {\lambda} C ^ {\mu} h ^ {\alpha \lambda} \right. \nonumber \\
& \left. - C ^ {\lambda} \partial _ {\lambda} h ^ {\alpha \mu} - \partial _ {\lambda} C ^ {\lambda} h ^ {\alpha \mu}\right) ) \delta \lambda \nonumber \\
& +  \partial _ {\lambda} C ^ {\alpha} C ^ {\lambda} \partial _ {\alpha} h ^ {\mu \nu} \delta \lambda \nonumber \\
& -  \partial _ {\alpha} C ^ {\alpha} (\partial ^ {\mu} C ^ {\nu} + \partial ^ {\nu} C ^ {\mu} - \eta ^ {\mu \nu} \partial _ {\lambda} C ^ {\lambda} +  \left(\partial _ {\alpha} C ^ {\mu} h ^ {\alpha \nu} +  \partial _ {\alpha} C ^ {\nu} h ^ {\alpha \mu} \right. \nonumber \\
&  \left. - C ^ {\lambda} \partial _ {\lambda} h ^ {\mu \nu} - \partial _ {\lambda} C ^ {\lambda} h ^ {\mu \nu}\right) ) \delta \lambda \nonumber \\
& -  (\partial ^ {\mu} \partial _ {\beta} C ^ {\nu} C ^ {\beta} + \partial _ {\beta} C ^ {\nu} \partial ^ {\mu} C ^ {\beta} ) \delta \lambda -  (\partial ^ {\nu} \partial _ {\beta} C ^ {\mu} C ^ {\beta} + \partial _ {\beta} C ^ {\mu} \partial ^ {\nu} C ^ {\beta} ) \delta \lambda \nonumber \\
&  + \eta ^ {\mu \nu}  (\partial _ {\alpha} \partial _ {\lambda} C ^ {\alpha} C ^ {\lambda} + \partial _ {\lambda} C ^ {\alpha} \partial _ {\alpha} C ^ {\lambda} ) \delta \lambda +  [-  (\partial _ {\alpha} \partial _ {\beta} C ^ {\nu} C ^ {\beta} + \partial _ {\beta} C ^ {\nu} \partial _ {\alpha} C ^ {\beta}) h ^ {\mu \alpha} \nonumber \\
& -  (\partial _ {\alpha} \partial _ {\beta} C ^ {\mu} C ^ {\beta} + \partial _ {\beta} C ^ {\mu} \partial _ {\alpha} C ^ {\beta}) h ^ {\alpha \nu} +  (\partial _ {\alpha} \partial _ {\lambda} C ^ {\alpha} C ^ {\lambda} + \partial _ {\lambda} C ^ {\alpha} \partial _ {\alpha} C ^ {\lambda}) h ^ {\mu \nu}] \delta \lambda \nonumber \\
& =  0 \,.
\end{align}
We see that all of the terms cancel due to the anticommutation of the partial derivatives of the ghost field. Moreover, the term $-  C ^ {\alpha} C ^ {\lambda} \partial _ {\alpha} \partial _ {\lambda} h ^ {\mu \nu} \delta \lambda$ vanishes due to the commutativity of partial derivatives.

\subsection{BRS Invariance of the Effective Action}

Assuming that $\partial _ {\nu}$ and $\delta _ {BRS}$ commute, then 
\be
\delta _ {BRS} F ^ {\tau} = \partial _ {\nu} \left(\delta _ {BRS} h ^ {\tau \nu} \right) \,. 
\ee
In addition,
\be
\delta _ {BRS} F _ {\tau} = \eta _ {\tau \mu} \partial _ {\nu} \left(\delta _ {BRS} h ^ {\mu \nu} \right) = \partial _ {\nu} \left(\delta _ {BRS} h _ {\tau} ^ {\nu}\right) \,.
\ee

In the variations to follow, we shall assume that $\delta _ {BRS}$ and $\cF(\Box)$ commute.

The variation of the gauge-fixing term is
\begin{align}
\delta _ {BRS} \left(- \frac{1}{2\xi}  \int{d ^ {4} x F _ {\tau} \cF(\Box) F ^ {\tau}} \right) & =  - \frac{1}{2\xi}  \int{d ^ {4} x \left(\delta _ {BRS} \left(F _ {\tau} \cF(\Box)  F ^ {\tau} \right)\right)} \nonumber \\
& =  - \frac{1}{2\xi}  \int{d ^ {4} x \left(\delta _ {BRS} \left(\eta _ {\tau \mu} \left(\partial _ {\nu} h ^ {\mu \nu}\right) \cF(\Box) F ^ {\tau} \right)\right)} \nonumber \\
& =  - \frac{1}{2\xi}  \int{d ^ {4} x \left(\delta _ {BRS} \left(\left(\partial _ {\nu} h ^ {\mu \nu}\right) \cF(\Box) F _ {\mu} \right)\right)} \nonumber \\
& =  - \frac{1}{2\xi}  \int{d ^ {4} x \left(\partial _ {\nu} \left(D _ {\alpha} ^ {\mu \nu} C ^ {\alpha} \delta \lambda \right) \cF(\Box) F _ {\mu}\right)} \nonumber \\ 
& - \frac{1}{2\xi}  \int{d ^ {4} x \left(\left(\partial _ {\nu} h ^ {\mu \nu}\right) \delta _ {BRS} \left(\cF(\Box) F _ {\mu} \right)\right)} \nonumber \\
& =  - \frac{1}{2\xi}  \int{d ^ {4} x \left(\partial _ {\nu} \left(D _ {\alpha} ^ {\mu \nu} C ^ {\alpha} \delta \lambda \right) \cF(\Box) F _ {\mu}\right)} \nonumber \\
& - \frac{1}{2\xi}  \int{d ^ {4} x \left(\left(\partial _ {\nu} h ^ {\mu \nu}\right) \cF(\Box) \left(\delta _ {BRS} F _ {\mu} \right)\right)} \,.
\end{align}
We know that $\Box = \partial _ {\mu} \partial ^ {\mu}$. Therefore, by integrating by parts successively (in fact, $2n$ times) in the second term and using the fact that the integrals of total divergences are equal to zero, we see that the second term is equal to the first term.

Hence,
\be
\delta _ {BRS} \left(- \frac{1}{2\xi}  \int{d ^ {4} x F _ {\tau} \cF(\Box)  F ^ {\tau}} \right) = - \frac{1}{\xi} \int{d ^ {4} x \left(\cF(\Box)  F _ {\tau} \left(\partial _ {\nu} \left(D _ {\alpha} ^ {\tau \nu} C ^ {\alpha} \right) \right) \delta \lambda \right)} \,.
\ee

The variation of the ghost-antighost action is
\begin{align}
\delta _ {BRS} \int{d ^ {4} x \left(\bar{C} _ {\tau} \overrightarrow{F} _ {\mu \nu} ^ {\tau} D _ {\alpha} ^ {\mu \nu} C ^ {\alpha} \right)} & =  \int{d ^ {4} x \left(\left(\delta _ {BRS} \bar{C} _ {\tau}\right) \overrightarrow{F} _ {\mu \nu} ^ {\tau} D _ {\alpha} ^ {\mu \nu} C ^ {\alpha} + \bar{C} _ {\tau} \delta _ {BRS} \left(\overrightarrow{F} _ {\mu \nu} ^ {\tau} D _ {\alpha} ^ {\mu \nu} C ^ {\alpha} \right)\right)} \nonumber \\
& =  - \frac{1}{\xi} \int{d ^ {4} x \left(\cF(\Box)  F _ {\tau} \delta \lambda \left(\delta _ {\mu} ^ {\tau} \partial _ {\nu} \left(D _ {\alpha} ^ {\mu \nu} C ^ {\alpha} \right) \right) + \bar{C} _ {\tau} \partial _ {\nu} \left(\delta _ {BRS} \left(D _ {\alpha} ^ {\tau \nu} C ^ {\alpha} \right) \right)\right)} \nonumber \\
& =  \frac{1}{\xi} \int{d ^ {4} x \left(\cF(\Box)  F _ {\tau} \left(\partial _ {\nu} \left(D _ {\alpha} ^ {\tau \nu} C ^ {\alpha} \right) \right) \delta \lambda \right)} \,,
\end{align}
due to the nilpotence of $h ^ {\mu \nu}$; we have also used the fact that $C ^ {\alpha}$ and $\delta \lambda$ anticommute. 

We observe that the variations of the gauge-fixing term and the ghost-antighost action cancel each other.

It can be seen that the computation of the variations does not depend on the power $n$ of the box operator in the gauge-fixing term and the BRS transformation of the antighost. Hence, the above results hold for any power $n$, so, by the principle of linear superposition, for a gauge-fixing term of the form
\be
- \frac{1}{2\xi}  \int{d ^ {4} x \, F _ {\tau} \cF(\Box) F ^ {\tau}}
\ee
and a BRS antighost transformation
\be
\delta _ {BRS} \bar{C} _ {\tau} = - \frac{1}{\xi} \cF(\Box) F _ {\tau} \delta \lambda \,.
\ee
Essentially, the only part of the ghost action which varies under the BRS transformations is the antighost $\bar{C} _ {\tau}$. Consequently, the BRS transformation of the antighost has been chosen to make the variation of the ghost action cancel the variation of the gauge-fixing term. Hence, the action $S_{\mt{quantised}}$ is BRS-invariant.

If we also include BRS-invariant couplings of the ghosts and gravitons to some external fields $K_{\mu \nu}$ (anti-commuting) and $L_{\sigma}$ (commuting),
we obtain the effective action $\widetilde{S}$, as:
\be
\widetilde{S} = S_{\mt{quantized}}+ K _ {\mu \nu} D _ {\al} ^ {\mu \nu} C ^ {\al} + L _ {\sigma} \p _ {\beta} C ^ {\sigma} C ^ {\bt}\,,
\ee
where $\widetilde{S}$ is also BRS-invariant.

\section{Slavnov identities}
\numberwithin{equation}{section}
\label{slav}

\subsection{Green's Functions}

Let us consider the expanded generating functional of Green's functions,
\begin{align}
Z[T _ {\mu \nu}, \bar{\beta} _ {\sigma}, \beta ^ {\tau}, K _ {\mu \nu}, L _ {\sigma}] &= N \int \left[\Pi _ {\mu \leq \nu} d h ^ {\mu \nu} \right]\left[d C ^ {\sigma}\right]\left[d \bar{C} _ {\tau}\right] \mathrm{exp} [i (\tilde{\Sigma} \left(h ^ {\mu \nu}, C ^ {\sigma}, \bar{C} _ {\tau}, K _ {\mu \nu}, L _ {\sigma} \right) \non 
&+ \bar{\beta} _ {\sigma} C ^ {\sigma} + \bar{C} _ {\tau} \beta ^ {\tau} + T _ {\mu \nu} h ^ {\mu \nu})] \,.
\end{align}
$\bar{\beta} _ {\sigma}$ and $\beta ^ {\tau}$ are anticommuting sources for the ghost and antighost fields respectively, while $K _ {\mu \nu}$ is an anticommuting external field and $L _ {\sigma}$ a commuting one.

The effective action is
\ba
\tilde{\Sigma} = S - \frac{1}{2\xi}   F _ {\tau} \cF (\Box)  F ^ {\tau} + \bar{C} _ {\tau} \overrightarrow{F} _ {\mu \nu} ^ {\tau} D _ {\alpha} ^ {\mu \nu} C ^ {\alpha} + K _ {\mu \nu} D _ {\alpha} ^ {\mu \nu} C ^ {\alpha} + L _ {\sigma} \partial _ {\beta} C ^ {\sigma} C ^ {\beta} \,.
\ea

We have that 
\be
\delta _ {BRS} \tilde{\Sigma} = 0 \,.
\ee 
Then
\begin{align}
\delta _ {BRS} h ^ {\mu \nu} \frac{\delta \tilde{\Sigma}}{\delta h ^ {\mu \nu}} + \delta _ {BRS} C ^ {\sigma} \frac{\delta \tilde{\Sigma}}{\delta C ^ {\sigma}} + \delta _ {BRS} \bar{C} _ {\tau} \frac{\delta \tilde{\Sigma}}{\delta \bar{C} _ {\tau}} & = 0 \non
\Rightarrow  D _ {\alpha} ^ {\mu \nu} C ^ {\alpha} \delta \lambda \frac{\delta \tilde{\Sigma}}{\delta h ^ {\mu \nu}} -  \partial _ {\beta} C ^ {\sigma} C ^ {\beta} \delta \lambda \frac{\delta \tilde{\Sigma}}{\delta C ^ {\sigma}} - \frac{1}{\xi} \cF(\Box)  F _ {\tau} \delta \lambda \frac{\delta \tilde{\Sigma}}{\delta \bar{C _ {\tau}}} &= 0 \non
\Rightarrow \frac{\delta \tilde{\Sigma}}{\delta K _ {\mu \nu}} \frac{\delta \tilde{\Sigma}}{\delta h ^ {\mu \nu}} + \frac{\delta \tilde{\Sigma}}{\delta L _ {\sigma}} \frac{\delta \tilde{\Sigma}}{\delta C ^ {\sigma}} + \frac{1}{\xi}\cF(\Box) F _ {\tau} \frac{\delta \tilde{\Sigma}}{\delta \bar{C _ {\tau}}} & = 0,
\end{align}
making use of the couplings. It should be pointed out that left variational derivatives are used with respect to anticommuting quantities:
\be
\delta f (C ^ {\sigma}) = \delta C ^ {\tau} \delta f / \delta C ^ {\tau} \,.
\ee

The reduced effective action is 
\be
\Sigma = \tilde{\Sigma} + \frac{1}{2\xi}  F _ {\tau} \cF(\Box)  F ^ {\tau} \,.
\ee

Since
\be
\frac{\delta \tilde{\Sigma}}{\delta K _ {\mu \nu}} = \frac{\delta \Sigma}{\delta K _ {\mu \nu}}, \frac{\delta \tilde{\Sigma}}{\delta L _ {\sigma}} = \frac{\delta \Sigma}{\delta L _ {\sigma}}, \frac{\delta \tilde{\Sigma}}{\delta C ^ {\sigma}} = \frac{\delta \Sigma}{\delta C ^ {\sigma}}, \frac{\delta \tilde{\Sigma}}{\delta \bar{C} _ {\tau}} = \frac{\delta \Sigma}{\delta \bar{C} _ {\tau}} \,,
\ee
we have
\begin{align}
& \frac{\delta \tilde{\Sigma}}{\delta K _ {\mu \nu}} \frac{\delta \tilde{\Sigma}}{\delta h ^ {\mu \nu}} + \frac{\delta \tilde{\Sigma}}{\delta L _ {\sigma}} \frac{\delta \tilde{\Sigma}}{\delta C ^ {\sigma}} + \frac{1}{\xi} \cF(\Box)  \overrightarrow{F} _ {\tau \mu \nu} h ^ {\mu \nu} \frac{\delta \tilde{\Sigma}}{\delta \bar{C _ {\tau}}} \nonumber \\
= & \frac{\delta \Sigma}{\delta K _ {\mu \nu}} \left(\frac{\delta \Sigma}{\delta h ^ {\mu \nu}} - \frac{\delta}{\delta h ^ {\mu \nu}} \left(\frac{1}{2\xi}  \left(\overrightarrow{F} _ {\tau \beta \gamma} h ^{\beta \gamma} \right) \cF(\Box)  \left(\overrightarrow{F} _ {\rho \sigma} ^ {\tau} h ^ {\rho \sigma} \right) \right) \right) \nonumber \\
&+ \frac{\delta \Sigma}{\delta L _ {\sigma}} \frac{\delta \Sigma}{\delta C ^ {\sigma}} + \frac{1}{\xi} \cF(\Box) \overrightarrow{F} _ {\tau \mu \nu} h ^ {\mu \nu} \frac{\delta \Sigma}{\delta \bar{C _ {\tau}}}.
\end{align}
But
\begin{align}
& \frac{\delta}{\delta h ^ {\mu \nu}} \left(\frac{1}{2\xi}  \left(\eta _ {\tau \beta} \partial _ {\gamma} h ^ {\beta \gamma} \right) \cF(\Box)  \left(\delta _ {\rho} ^ {\tau} \partial _ {\sigma} h ^ {\rho \sigma} \right) \right) \nonumber \\
= & \frac{1}{2\xi}  \eta _ {\tau \beta} \delta _ {\rho} ^ {\tau} \frac{\delta}{\delta h ^ {\mu \nu}} \left[\left(\partial _ {\gamma} h ^ {\beta \gamma} \right) \cF(\Box) \left(\partial _ {\sigma} h ^ {\rho \sigma} \right) \right] \nonumber \\
= & - \frac{1}{2\xi}  \eta _ {\rho \beta} \left[\frac{\delta h ^ {\beta \gamma}}{\delta h ^ {\mu \nu}} \partial _ {\gamma} \left(\cF(\Box)  \left(\partial _ {\sigma} h ^ {\rho \sigma} \right) \right) + \frac{\delta h ^ {\rho \sigma}}{\delta h ^ {\mu \nu}} \partial _ {\sigma} \left(\cF(\Box)  \left(\partial _ {\gamma} h ^{\beta \gamma} \right) \right) \right] \,,
\end{align}
after applying integration by parts before functional differentiation once in the first term and $2n + 1$ times in the second term and using the Leibniz rule. We see that the first term is equal to the second term, thereby eliminating the $\frac{1}{2}$ factor. Performing integration by parts one final time, we obtain the relation
\be
\frac{\delta \Sigma}{\delta K _ {\mu \nu}}\frac{\delta \Sigma}{\delta h ^ {\mu \nu}} + \frac{\delta \Sigma}{\delta L _ {\sigma}}\frac{\delta \Sigma}{\delta C ^ {\sigma}} = 0 \,.
\ee

We also have that 
\begin{align}
 \overrightarrow{F} _ {\mu \nu} ^ {\tau} \frac{\delta \Sigma}{\delta K _ {\mu \nu}} & =   \overrightarrow{F} _ {\mu \nu} ^ {\tau} \frac{\delta \tilde{\Sigma}}{\delta K _ {\mu \nu}} \nonumber \\
& =   \delta _ {\mu} ^ {\tau} \partial _ {\nu} \frac{\delta \tilde{\Sigma}}{\delta K _ {\mu \nu}} \nonumber \\
& =   \partial _ {\nu} \frac{\delta \tilde{\Sigma}}{\delta K _ {\tau \nu}} \nonumber \\
& =  \partial _ {\nu} \left(D _ {\alpha} ^ {\tau \nu} C ^ {\alpha} \right) \nonumber \\
& = \frac{\delta \tilde{\Sigma}}{\delta \bar{C _ {\tau}}} \nonumber \\
& =  \frac{\delta \Sigma}{\delta \bar{C _ {\tau}}}\,.
\end{align}
 
To prove that the metric
\be
\left[\Pi _ {\mu \leq \nu} d h ^ {\mu \nu}\right]\left[d C ^ {\sigma}\right]\left[d \bar{C} _ {\tau}\right]
\ee
is BRS invariant, we must show that the corresponding Jacobian, $\mathrm{det} J$, is equal to $1$.
 
We have that
\be
\int{d ^ 4 x \partial _ {\alpha} C ^ {\alpha}} = 0
\ee
by the divergence theorem. Then
\begin{align}
\frac{\delta ^ 2 \tilde{\Sigma}}{\delta K _ {(\mu \nu)} \delta h ^ {(\mu \nu)}} & =  \frac{\delta}{\delta h ^ {(\mu \nu)}} \left( D _ {\alpha} ^ {(\mu \nu)} C ^ {\alpha} \right) \nonumber \\
& =  \frac{\delta}{\delta h ^ {(\mu \nu)}} \left( \left(\partial _ {\alpha} C ^ {(\mu \mid} h ^ {\alpha \mid \nu)} +  \partial _ {\alpha} C ^ {(\nu \mid} h ^ {\alpha \mid \mu)} - C ^ {\alpha} \partial _ {\alpha} h ^ {(\mu \nu)} - \partial _ {\alpha} C ^ {\alpha} h ^ {(\mu \nu)}\right)\right) \nonumber \\
& =  0 \,, \\
\frac{\delta ^ 2 \tilde{\Sigma}}{\delta C ^ {\sigma} \delta L _ {\sigma}} & =  \frac{\delta}{\delta C ^ {\sigma}} \left( \partial _ {\beta} C ^ {\sigma} C ^ {\beta} \right) \nonumber \\
& =  \frac{\delta}{\delta C ^ {\sigma}} \left(-  C ^ {\beta} \partial _ {\beta} C ^ {\sigma} \right) \nonumber \\
& =  -  \frac{\delta C ^ {\beta}}{\delta C ^ {\sigma}} \partial _ {\beta} C ^ {\sigma} -  \partial _ {\beta} C ^ {\beta} \nonumber \\
& = 0 \,,
\end{align}
by the relation above, the Leibniz rule, the anticommuting property of the ghost field and by integrating by parts where appropriate. The parentheses surrounding the indices indicate that the summation is to be carried out only for $\mu \leq \nu$.

Then
\begin{align}
\mathrm{det} J & =  \left| \begin{array}{ccc}
\frac{\delta \left(h ^ {\mu \nu} + \delta _ {\mathrm{BRS}} h ^ {\mu \nu} \right)}{\delta h ^ {\alpha \beta}} & \frac{\delta \left(h ^ {\mu \nu} + \delta _ {\mathrm{BRS}} h ^ {\mu \nu} \right)}{\delta C ^ {\gamma}} & 0 \\
0 & \frac{\delta \left(C ^ {\sigma} + \delta _ {\mathrm{BRS}} C ^ {\sigma} \right)}{\delta C ^ {\gamma}} & 0 \\
\frac{\delta \left(\bar{C} _ {\tau} + \delta _ {\mathrm{BRS}} \bar{C} _ {\tau} \right)}{\delta h ^ {\alpha \beta}} & 0 & \frac{\delta \left(\bar{C} _ {\tau} + \delta _ {\mathrm{BRS}} \bar{C} _ {\tau} \right)}{\delta \bar{C} _ {\delta}} 
\end{array} \right| \nonumber \\
& = \mathrm{det} (I + A) \nonumber \\
& =  \exp (\mathrm{Tr} (\log (I + A))) \nonumber \\
& =  1 + \mathrm{Tr} A + \mathcal{O} (A ^ 2) \nonumber \\
& =  1 + \left(\frac{\delta ^ 2 \tilde{\Sigma}}{\delta K _ {(\mu \nu)} \delta h ^ {(\mu \nu)}} + \frac{\delta ^ 2 \tilde{\Sigma}}{\delta C ^ {\sigma} \delta L _ {\sigma}} \right)\delta \lambda \nonumber \\
& =  1 \,,
\end{align}
using the relations above and employing the series expansion of the $\log (1+x)$ function. We have also used the fact that $\delta \lambda ^ 2 = 0$.

Keeping in mind that $\delta _ {BRS} \tilde{\Sigma} = 0$ and performing the BRS variation of $\tilde{\Sigma} + \kappa T _ {\mu \nu} h ^ {\mu \nu} + \bar{\beta _ {\sigma}} C ^ {\sigma} + \bar{C _ {\tau}} \beta ^ {\tau}$ with respect to $h ^ {\mu \nu}$, $C ^ {\sigma}$ and $\bar{C _ {\tau}}$, we obtain
\begin{align}
& N \int\left[\Pi _ {\mu \leq \nu} d h ^ {\mu \nu} \right]\left[d C ^ {\sigma}\right]\left[d \bar{C} _ {\tau}\right] \left( T _ {\mu \nu} D _ {\alpha} ^ {\mu \nu} C ^ {\alpha} -  \bar{\beta} _ {\sigma} \partial _ {\beta} C ^ {\sigma} C ^ {\beta} + \frac{1}{\xi} \beta ^ {\tau} \cF(\Box)  \overrightarrow{F} _ {\tau \mu \nu} h ^ {\mu \nu} \right) \non 
& \times \exp \left[i \left(\tilde{\Sigma} +  T _ {\mu \nu} h ^ {\mu \nu} + \bar{\beta} _ {\sigma} C ^ {\sigma} + \bar{C} _ {\tau} \beta ^ {\tau} \right)\right] = 0 \,,
\end{align}
where we have expanded up to first order in the generating functional ($\delta \lambda ^ 2 = 0$). Changing variables does not change the value of the generating functional, while the measure does not change too. It is assumed that $\delta \lambda$ anticommutes with $\bar{\beta} _ {\sigma}$ and $\beta ^ {\tau}$.

By varying $\tilde{\Sigma} +  T _ {\mu \nu} h ^ {\mu \nu} + \bar{\beta _ {\sigma}} C ^ {\sigma} + \bar{C _ {\tau}} \beta ^ {\tau}$ with respect to $\bar{C _ {\tau}} \rightarrow \bar{C _ {\tau}} + \delta \bar{C _ {\tau}}$, we get the ghost equation of motion:
\be
N \int{\left[\Pi _ {\mu \leq \nu} d h ^ {\mu \nu} \right]\left[d C ^ {\sigma}\right]\left[d \bar{C} _ {\tau}\right] \left(\frac{\delta \tilde{\Sigma}}{\delta \bar{C} _ {\tau}} + \beta ^ {\tau} \right) \mathrm{exp} \left[i \left(\tilde{\Sigma} +  T _ {\mu \nu} h ^ {\mu \nu} + \bar{\beta} _ {\sigma} C ^ {\sigma} + \bar{C} _ {\tau} \beta ^ {\tau} \right)\right]} = 0 \,;
\ee
this equation comes from expanding the generating functional up to first order ($\delta \bar{C} _ {\tau} ^ 2 = 0$) and writing the variation $\delta \tilde{\Sigma} + \delta \bar{C} _ {\tau} \beta ^ {\tau}$ as $\delta \bar{C} _ {\tau} \left(\frac{\delta \tilde{\Sigma}}{\delta \bar{C} _ {\tau}} + \beta ^ {\tau} \right)$ ($\delta \bar{C} _ {\tau}$ is arbitrary so it cancels). Shift invariance being the defining property of the Grassmann integral, the value of the generating functional does not change. The measure also stays the same.

The generating functional of connected Green's functions is
\be
W[T _ {\mu \nu}, \bar{\beta} _ {\sigma}, \beta ^ {\tau}, K _ {\mu \nu}, L _ {\sigma}] = -i \log Z[T _ {\mu \nu}, \bar{\beta} _ {\sigma}, \beta ^ {\tau}, K _ {\mu \nu}, L _ {\sigma}] \,.
\ee

We observe that 
\be
 T _ {\mu \nu} \frac{\delta Z}{\delta K _ {\mu \nu}} - \bar{\beta _ {\sigma}} \frac{\delta Z}{\delta L _ {\sigma}} + \frac{1}{\xi} \beta ^ {\tau} \cF(\Box)  \overrightarrow{F} _ {\tau \mu \nu} \frac{\delta Z}{\delta T _ {\mu \nu}} = 0 \,, 
\ee
making use of the couplings. The same equation holds for $W$ in the place of $Z$. 

By the same argument as above, we obtain the ghost equation of motion:
\be
\kappa ^ {-1} \overrightarrow{F} _ {\mu \nu} ^ {\tau} \frac{\delta W}{\delta K _ {\mu \nu}} + \beta ^ {\tau} = 0 \,.
\ee

\subsection{Proper Vertices}

The expectation values of the gravitational, ghost, and antighost fields are
\begin{align}
<h ^ {\mu \nu} (x)> & = \frac{\delta W}{\kappa \delta T _ {\mu \nu} (x)} \,, \\
<C ^ {\sigma} (x)> & = \frac{\delta W}{\delta \bar{\beta} _ {\sigma} (x)} \,,\\
<\bar{C} _ {\tau} (x)> & = - \frac{\delta W}{\delta \beta ^ {\tau} (x)} \,.
\end{align}
We have assumed that the ghost field anticommutes with $\bar{\beta} _ {\sigma}$ and that the antighost field anticommutes with $\beta ^ {\tau}$.

The generating functional of proper vertices is
\be
\tilde{\Gamma} [h ^ {\mu \nu}, C ^ {\sigma}, \bar{C} _ {\tau}, K _ {\mu \nu}, L _ {\sigma} ] = W [T _ {\mu \nu}, \bar{\beta} _ {\sigma}, \beta ^ {\tau}, K _ {\mu \nu}, L _ {\sigma} ] -  T _ {\mu \nu} h ^ {\mu \nu} - \bar{\beta} _ {\sigma} C ^ {\sigma} - \bar{C} _ {\tau} \beta ^ {\tau} \,.
\ee

The dual relations are
\begin{align}
 T _ {\mu \nu} (x) & = - \frac{\delta \tilde{\Gamma}}{\delta <h ^ {\mu \nu} (x)>} \,, \\
\bar{\beta} _ {\sigma} (x) & = \frac{\delta \tilde{\Gamma}}{\delta <C ^ {\sigma} (x)>} \,, \\
\beta ^ {\tau} (x) & = - \frac{\delta \tilde{\Gamma}}{\delta <\bar{C} _ {\tau} (x)>} \,.
\end{align}

We also have that
\begin{align}
\frac{\delta \tilde{\Gamma}}{\delta K _ {\mu \nu} (x)} & = \frac{\delta W}{\delta K _ {\mu \nu} (x)} \,,\\
\frac{\delta \tilde{\Gamma}}{\delta L _ {\sigma} (x)} & = \frac{\delta W}{\delta L _ {\sigma} (x)} \,. 
\end{align}

Obtaining the relations
\be
\frac{\delta \tilde{\Gamma}}{\delta K _ {\mu \nu}} \frac{\delta \tilde{\Gamma}}{\delta h ^ {\mu \nu}} + \frac{\delta \tilde{\Gamma}}{\delta L _ {\sigma}} \frac{\delta \tilde{\Gamma}}{\delta C ^ {\sigma}} + \frac{1}{\xi} \cF(\Box)  \overrightarrow{F} _ {\tau \mu \nu} h ^ {\mu \nu} \frac{\delta \tilde{\Gamma}}{\delta \bar{C} _ {\tau}} = 0
\ee
and
\be
 \overrightarrow{F} _ {\mu \nu} ^ {\tau} \frac{\delta \tilde{\Gamma}}{\delta K _ {\mu \nu}} - \frac{\delta \tilde{\Gamma}}{\delta \bar{C} _ {\tau}} = 0
\ee
is trivial.

Let us now define a reduced generating functional of proper vertices,
\be
\Gamma = \tilde{\Gamma} + \frac{1}{2\xi}  \left(\overrightarrow{F} _ {\tau \mu \nu} h ^ {\mu \nu} \right) \cF(\Box)  \left(\overrightarrow{F} _ {\rho \sigma} ^ {\tau} h ^ {\rho \sigma} \right) \,.
\ee

Following exactly the same argument as for the reduced effective action, we get the Slavnov identity
\be
\frac{\delta \Gamma}{\delta K _ {\mu \nu}} \frac{\delta \Gamma}{\delta h ^ {\mu \nu}} + \frac{\delta \Gamma}{\delta L _ {\sigma}} \frac{\delta \Gamma}{\delta C ^ {\sigma}} = 0
\ee
and the ghost equation of motion
\be
 \overrightarrow{F} _ {\mu \nu} ^ {\tau} \frac{\delta \Gamma}{\delta K _ {\mu \nu}} - \frac{\delta \Gamma}{\delta \bar{C} _ {\tau}} = 0 \,.
\ee

\subsection{Local Solutions}

Let us write $\mathcal{G}$ as
\be
\mathcal{G} = \mathcal{G} _ 1 + \mathcal{G} _ {0} \,,
\ee
where
\be
\mathcal{G} _ 1 = \frac{\delta \Sigma}{\delta h ^ {\mu \nu}} \frac{\delta}{\delta K _ {\mu \nu}} + \frac{\delta \Sigma}{\delta C ^ {\beta}} \frac{\delta}{\delta L _ {\beta}}
\ee
and
\be
\mathcal{G} _ 0 = \frac{\delta \Sigma}{\delta K _ {\rho \sigma}} \frac{\delta}{\delta h ^ {\rho \sigma}} + \frac{\delta \Sigma}{\delta L _ {\tau}} \frac{\delta}{\delta C ^ {\tau}} \,.
\ee

We have that $\mathcal{G} ^ {2} = \left(\mathcal{G} _ {1} + \mathcal{G} _ {0} \right) \left(\mathcal{G} _ {1} + \mathcal{G} _ {0} \right) = \mathcal{G} _ {1} ^ {2} + \mathcal{G} _ {1} \mathcal{G} _ {0} + \mathcal{G} _ {0} \mathcal{G} _ {1} + \mathcal{G} _ {0} ^2 = \mathcal{G} _ {1} ^ {2} + \mathcal{G} _ {1} \mathcal{G} _ {0} + \mathcal{G} _ {0} \mathcal{G} _ {1}$, since $\mathcal{G} _ {0}$ is nilpotent. 
If we act on the relation
\be
\frac{\delta \Sigma}{\delta K _ {\mu \nu}}\frac{\delta \Sigma}{\delta h ^ {\mu \nu}} + \frac{\delta \Sigma}{\delta L _ {\sigma}}\frac{\delta \Sigma}{\delta C ^ {\sigma}} = 0
\ee
with $\frac {\delta}{\delta C}$, $\frac {\delta}{\delta L}$, $\frac {\delta}{\delta K}$, $\frac {\delta}{\delta h}$ (indices are suppressed) where appropriate, then we get
\begin{align}
\mathcal{G} ^ {2} & = \left(\frac{\delta \Sigma}{\delta h ^ {\mu \nu}} \frac{\delta}{\delta K _ {\mu \nu}} + \frac{\delta \Sigma}{\delta C ^ {\beta}} \frac{\delta}{\delta L _ {\beta}}\right) \left(\frac{\delta \Sigma}{\delta h ^ {\rho \sigma}} \frac{\delta}{\delta K _ {\rho \sigma}} + \frac{\delta \Sigma}{\delta C ^ {\tau}} \frac{\delta}{\delta L _ {\tau}}\right) \nonumber \\ 
& + \left(\frac{\delta \Sigma}{\delta h ^ {\mu \nu}} \frac{\delta}{\delta K _ {\mu \nu}} + \frac{\delta \Sigma}{\delta C ^ {\beta}} \frac{\delta}{\delta L _ {\beta}}\right) \left(\frac{\delta \Sigma}{\delta K _ {\rho \sigma}} \frac{\delta}{\delta h ^ {\rho \sigma}} + \frac{\delta \Sigma}{\delta L _ {\tau}} \frac{\delta}{\delta C ^ {\tau}}\right) \nonumber \\
& + \left(\frac{\delta \Sigma}{\delta K _ {\rho \sigma}} \frac{\delta}{\delta h ^ {\rho \sigma}} + \frac{\delta \Sigma}{\delta L _ {\tau}} \frac{\delta}{\delta C ^ {\tau}}\right) \left(\frac{\delta \Sigma}{\delta h ^ {\mu \nu}} \frac{\delta}{\delta K _ {\mu \nu}} + \frac{\delta \Sigma}{\delta C ^ {\beta}} \frac{\delta}{\delta L _ {\beta}}\right) \nonumber \\
& = \frac{\delta \Sigma}{\delta h} \left(\frac{\delta ^ {2} \Sigma}{\delta K \delta h} \frac{\delta}{\delta K} + \frac{\delta \Sigma}{\delta h} \frac{\delta ^ {2}}{\delta K \delta K} + \frac{\delta ^ {2} \Sigma}{\delta K \delta C} \frac{\delta}{\delta L} + \frac{\delta \Sigma}{\delta C} \frac{\delta ^ {2}}{\delta K \delta L}  \right) \nonumber \\
& + \frac{\delta \Sigma}{\delta C} \left(\frac{\delta ^ {2} \Sigma}{\delta L \delta h} \frac{\delta}{\delta K} + \frac{\delta \Sigma}{\delta h} \frac{\delta ^ {2}}{\delta L \delta K} + \frac{\delta ^ {2} \Sigma}{\delta L \delta C} \frac{\delta}{\delta L} + \frac{\delta \Sigma}{\delta C} \frac{\delta ^ {2}}{\delta L \delta L}  \right) \nonumber \\
& + \frac{\delta \Sigma}{\delta h} \left(\frac{\delta ^ {2} \Sigma}{\delta K \delta K} \frac{\delta}{\delta h} + \frac{\delta \Sigma}{\delta K} \frac{\delta ^ {2}}{\delta K \delta h} + \frac{\delta ^ {2} \Sigma}{\delta K \delta L} \frac{\delta}{\delta C} + \frac{\delta \Sigma}{\delta L} \frac{\delta ^ {2}}{\delta K \delta C}  \right) \nonumber \\
& + \frac{\delta \Sigma}{\delta C} \left(\frac{\delta ^ {2} \Sigma}{\delta L \delta K} \frac{\delta}{\delta h} + \frac{\delta \Sigma}{\delta K} \frac{\delta ^ {2}}{\delta L \delta h} + \frac{\delta ^ {2} \Sigma}{\delta L \delta L} \frac{\delta}{\delta C} + \frac{\delta \Sigma}{\delta L} \frac{\delta ^ {2}}{\delta L \delta C}  \right) \nonumber \\
&+ \frac{\delta \Sigma}{\delta K} \left(\frac{\delta ^ {2} \Sigma}{\delta h \delta h} \frac{\delta}{\delta K} + \frac{\delta \Sigma}{\delta h} \frac{\delta ^ {2}}{\delta h \delta K} + \frac{\delta ^ {2} \Sigma}{\delta h \delta C} \frac{\delta}{\delta L} + \frac{\delta \Sigma}{\delta C} \frac{\delta ^ {2}}{\delta h \delta L}  \right) \nonumber \\
&+ \frac{\delta \Sigma}{\delta L} \left(\frac{\delta ^ {2} \Sigma}{\delta C \delta h} \frac{\delta}{\delta K} + \frac{\delta \Sigma}{\delta h} \frac{\delta ^ {2}}{\delta C \delta K} + \frac{\delta ^ {2} \Sigma}{\delta C \delta C} \frac{\delta}{\delta L} + \frac{\delta \Sigma}{\delta C} \frac{\delta ^ {2}}{\delta C \delta L}  \right) \nonumber \\
&= 0 \,.
\end{align}
The equation above leads us to consider local solutions to the renormalisation equation of the form
\be
\Gamma _ {\mathrm{div}} ^ {(n)} = \mathcal{S}(h ^ {\mu \nu}) + \mathcal{G}[X(h ^ {\mu \nu}, C ^ {\sigma}, \bar{C} _ {\tau}, K _ {\mu \nu}, L _ {\sigma})] \,,
\ee
where $\mathcal{S}$ is an arbitrary gauge-invariant local functional of $h ^ {\mu \nu}$ and its derivatives, and $X$ is an arbitrary local functional of $h ^ {\mu \nu}$, $C ^ {\sigma}$, $\bar{C} _ {\tau}$, $K _ {\mu \nu}$, and $L _ {\sigma}$ and their derivatives.

Now, to satisfy the ghost equation of motion, we require
\be
\Gamma _ {\mathrm{div}} ^ {(n)} = \Gamma _ {\mathrm{div}} ^ {(n)} (h ^ {\mu \nu}, C ^ {\sigma}, K _ {\mu \nu} - \kappa ^ {-1} \bar{C} _ {\tau} \overleftarrow{F} _ {\mu \nu} ^ {\tau}, L _ {\sigma}) \,. 
\ee
The relation above may be verified by substituting the said requirement back into the ghost equation of motion, using the chain rule and applying integration by parts.


\section{Conclusion and Outlook}
\numberwithin{equation}{section}
\label{sec:concl}

We have seen that, for $L>1$, the loop amplitudes for our infinite-derivative gravitational action are superficially convergent, which is a promising hint as far as renormalisability is concerned. Moreover, the infinite-derivative gravitational action is invariant under the BRS (Becchi-Rouet-Stora) transformations.

Moreover, we explicitly show that the corresponding Slavnov identities hold for our action. For a renormalisation program to be successful, this is a prerequisite. Obviously, the next step would be to compute Feynman diagrams for the theory and demonstrate that we do not get higher and higher divergences as the loop order increases. That is, a finite set of counterterms ensures renormalisability.
For an infinite-derivative scalar toy model~\cite{Biswas:2014tua,Talaganis:2014ida,Talaganis:2017tnr}, there have been compelling arguments that the toy model is UV finite, employing dressed propagators and dressed vertices. It remains to be seen whether the same line of argument can be carried out for our infinite-derivative gravitational action.

\section{Acknowledgements}

ST is supported by a scholarship from the Onassis Foundation.

\end{document}